\documentclass[runningheads]{llncs}
\usepackage[T1]{fontenc}

\usepackage{subcaption}
\usepackage{sectsty}
\usepackage{listings}
\lstdefinestyle{pythonstyle}{
  language=Python,
  basicstyle=\ttfamily\footnotesize,
  keywordstyle=\color{blue}\bfseries,
  numbers=left,
  stepnumber=1,
  numbersep=5pt,
  showspaces=false,
  showstringspaces=false,
  tabsize=4,
  breaklines=true,
  breakatwhitespace=true,
  keepspaces=true,
  columns=flexible,
  lineskip=-2pt
}

\usepackage{graphicx}
\PassOptionsToPackage{hyphens}{url}\usepackage{hyperref}

\usepackage{todonotes}
\setuptodonotes{inline}

\usepackage{color}

\begin{document}

\title{Quality evaluation of Tabby coding assistant using real source code snippets}
\titlerunning{Tabby coding assistant code quality evaluation}
%
\author{Marta Borek\orcidID{0009-0002-5538-6556}\email{marta.borek.stud@pw.edu.pl}\\
  and
  Robert Nowak\orcidID{0000-0001-7248-6888}\email{robert.nowak@pw.edu.pl}}
\authorrunning{M. Borek, R. Nowak}
%
\institute{Institute of Computer Science, Warsaw University of Technology}

\maketitle              
\begin{abstract}
 Large language models have become a popular tool in software development, providing coding assistance. The proper measurement of the accuracy and reliability of the code produced by such tools is a challenge due to natural language prompts.

  We propose a simple pipeline that uses state-of-the-art implementation
  of classic and universal genres of algorithms and data structures.
  We focus on measuring the quality of TabbyML code assistant due to its
  open licence and the flexibility in the choice of the language model.

  Our results presented as cyclomatic complexity, Halstead's Bugs \& Effort and four text-based similarity matrices depict the usability of TabbyML in coding assistance tasks.

\end{abstract}
\section{Introduction}
The widespread adoption of generative AI in software development has
introduced coding assistants, often referred to as ``copilots'', due
to the influence of GitHub Copilot, one of the precursors in the field.
Connected to integrated development environments (IDEs), these tools
work alongside programmers, offering benefits such as reduced typing
effort, minimized code repetitiveness, and promotion of good
programming practices by leveraging large language models (LLMs)
trained on big datasets of good quality code. What is more, they
enable developers to focus on high-level problem-solving while
automating the more basic implementation tasks.

However, despite these advantages, concerns arise regarding accuracy
and reliability of the code produced by coding assistants. While high
quality suggestions streamline development, incorrect ones can cause
confusion, leading either to a time-consuming refactoring process or
the need to continuously reject completions, both disrupting the
programming workflow.

Ensuring the quality of AI-generated code remains a major challenge,
as a term as broad as code quality lacks a single objective metric.
Factors such as coding standards and style, demand for readability,
efficiency, and execution speed vary across projects, making the
evaluation of potential suggestions highly subjective. While
adjustable parameters of some coding assistants---context scope and
choice of specific model---help optimize accuracy, a perception of
correctness remains inherently personal, highlighting the complexity
of assessing coding assistant's performance in generating completions.

A number of research was released in recent months, occupied with the
subject of generative AI quality evaluation. Two studies were
particularly influential to the direction of this work. The first one
utilized OpenAI's \textit{HumanEval} dataset
\cite{chen2021evaluating}, comprised of carefully curated
problem-solving test cases, not present in the model's training data
and the second one focused on evaluating AI-generated Python code
\cite{yang2024evaluation}.
Both studies take into consideration prominent LLM-powered tools:
Github Copilot, ChatGPT, Microsoft Copilot, Tabnine and Codeium,
generating completions through a set of prompts and comparing them in
a dedicated testing environment.

Given the extent of these studies, they deliver a great point of
reference and a lot of insightful data.
Test scenarios evaluate each individual assistant with an exhaustive
use of metrics such as average lines of code, McCabe complexity, code
generation time, runtime efficiency, memory usage, error handling and
most importantly, test-case success rate.
However, while the test cases focus on individual assessments, one
gap in these studies lies in their overall comparative nature.
Upon benchmarking a number of tools, the outcomes are presented in a
relative manner making it possible to state that some tools fare
better than others in given conditions.
Nevertheless, for establishing the quality of a single tool, as a
standalone, the discussed testing environments lack definite point of reference.

What is more, the way that code snippets from coding assistants are
obtained in these studies is through natural language prompts,
which---while an NL-driven built-in chat is usually one of the
features typically offered by coding assistants---fails to capture
the organic way of serving completions during the process of coding,
based solely on the code scope of context.

TabbyML is an open-source coding assistant, having the flexibility in
the choice of the language model, allowing for great scalability for
different tasks and platforms, enabling even consumer-grade GPUs to
run the service.
Its especially powerful characteristic is the self hosting ability,
which offers an undisputable advantage of granting users complete
independence from external providers. This makes Tabby a remarkably
robust choice from the privacy standpoint, as it eliminates any
telemetry and data mining risks, a controversy often associated with
this niche of tools.

This work addresses the gap regarding the environment for prompt
generation, with the effort to bring it closer to natural conditions of
coding assistant employment. The section methodology depicts the new pipeline
used for testing, the section results included the metrices for TabbyML,
and last section concludes our work.

\section{Methodology}

The subject of our experiments is a single coding assistant,
Tabby, which creates the demand for the much emphasized point of reference.
This element was established with the use of the source set from the
Algorithms database and code clone and similarity detection
techniques---methodology adapted for this work.

\begin{figure*}[!htb]
  \centering
  \includegraphics[width=\textwidth]{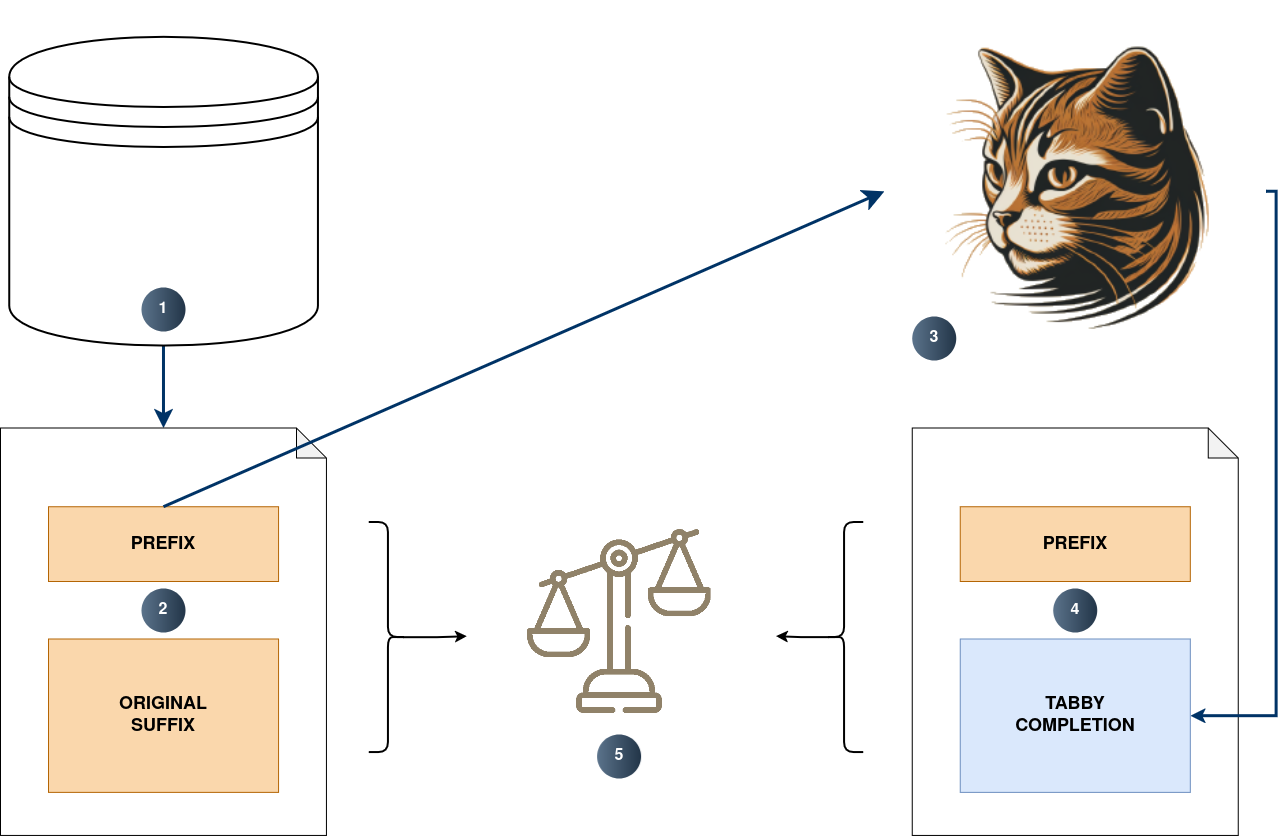}
  \caption[Testing pipeline.]{Diagram of the testing pipeline's
    architecture; (1) Data Preprocessing and Selection, (2) Prefix
    Selection, (3) Server Interaction, (4) Tabby Completed Program, (5) Similarity Testing.}
  \label{fig:pipeline}
\end{figure*}

This novel approach assumes the original implementations from the
dataset as point of reference, trusting in their community-verified quality.
Tabby-completed programs, treated as potential clones are compared
against the originals, in search for similarity.
Distinct steps of the testing process are depicted in Fig.~\ref{fig:pipeline}.

\subsection{Data Preprocessing and Selection}

The Algorithms \cite{thealgorithms} database, an open Github repository with over 197k stars,
was selected as ground truth data. This choice was dictated mainly by narrowing
coding tasks to a classic and universal genre of algorithms and data structures.
The Algorithms database is a collection of algorithms' implementations for distinct programming languages.
The Python subset was used in our study.
Python modules of The Algorithms database are focused on: data structures, digital
image processing, divide and conquer, dynamic programming, fractals,
graphs, greedy methods, hashes, maths, scheduling, searches, sorts
and web programming algorithms were selected as a source set for
prompt generation.

\subsection{Prompt Prefix Selection}

The most influential element in generating code suggestions is the
choice of prefix---a working file's excerpt preceding the completion
invocation point.
Prompt base of prefixes was obtained from the Algorithms subset's
source files as increasingly larger parts, starting with 10\% of the
file's content and incrementing by a 10\% points step with each
iteration, yielding 9 prefix cases for one original sample---1st and
2nd steps of the figure's \ref{fig:pipeline} diagram.

The randomization in contents and prefix termination points
introduced with this method captures the nature of real-world
interaction with coding assistants, as completions may be invoked
mid-word, having only the limited amount of preceding code as a
source of context.

\subsection{Server Interaction}

The Tabby server ran on a remote machine equipped with an Nvidia RTX
A5000 GPU and 60GB RAM, powered by the StarCoder-1B model. While
definitely the smallest in scale option from Tabby's repository, it
was chosen for its accessibility for virtually any consument-grade
hardware platform. This was done to ensure that a regular user,
unable to host more complex models locally, could still benefit from
the tool. An assumption that performance improvements would scale
with larger models played a significant part in that decision as well.

For each prefix obtained in the prefix selection step, an HTTP
request was sent to Tabby instance---3rd stage of the diagram in
Fig.~\ref{fig:pipeline}---in the required request body, displayed
in listing~\ref{lst:tabby_prompt}.
Python was specified as programming language for all cases and the
suffix was ommited to simplify the methodology.

\begin{lstlisting}[caption={Tabby's prompt with an example fibonacci prefix and suffix, from Tabby's Swagger API},
  label=lst:tabby_prompt]
{
  "language": "python",
  "segments": {
    "prefix": "def fib(n):\n    ",
    "suffix": "\n        return fib(n - 1) + fib(n - 2)"
  }
}
\end{lstlisting}

\subsection{Tabby Completed program}

Each Tabby-generated suggestion was merged with its original prefix
to form a variant to the original program---diagram step 4 in figure
\ref{fig:pipeline}---and enable the code clone detection-based
testing methodology.

\subsection{Similarity metrics}

\paragraph{Static metrics}

Static metrics offer a cost-effective way to derive objective,
quantifiable insights, providing a solid foundation for evaluating
code quality. This experiment included two key static metrics:
\begin{itemize}
  \item \textbf{Cyclomatic Complexity}, which measures the number of
    independent execution paths in the code.
  \item \textbf{Halstead's Bugs \& Effort} to estimate potential
    vulnerabilities and cognitive effort required for code
    comprehension or implementation.
\end{itemize}

These metrics were chosen for their ability to capture fundamental
code characteristics, making them well-suited for the relatively
simple codebase in this study.
The experiment utilized Radon, a Python tool for code analysis, to
compute metric values for each original program and its duplicates.
Additionally, the length ratio of duplicate files to their originals
was calculated, providing valuable insight for both aspects of the analysis.

\paragraph{Similarity detection}

A key stage in the process is measuring how closely Tabby-generated
suggestions match the original code. Code clone detection methods
vary based on the types of clones they identify and their outlook and
approach to code in analysis. This experiment used text-based
algorithms for simplicity reasons, as they evaluate code purely by
its textual properties, treating programs as character sequences.
Undeniable benefits of this approach include computational
efficiency, analyzed programming language flexibility, and minimal
subjects' preprocessing. However, one major limitation is the
inability to detect functional clones---programs with identical
behavior but different structure---since identifying such duplicates
requires a test-based environment to analyze the code execution
process, which is far more complex and resource-intensive.

Four algorithms falling into this text-based category were selected, notably:
\begin{itemize}
  \item \textbf{Ratcliff-Obershelp} operates on the basis of
    sequences. Its score is a [0, 1] value calculated by dividing all
    matching characters by the combined lengths of both strings. It
    is known for its accurate reflections of the subjective human
    estimation of similarity.
  \item \textbf{Jaro-Winkler similarity} is a distance-based
    algorithm, building upon \textit{Jaro similarity} formula, by
    assigning greater importance to characters matching from the
    beginning of sequences within a common prefix length.
  \item \textbf{Hamming distance} is calculated as a number of
    locations between two strings where the corresponding characters
    differ. From the edit-based standpoint, it is the number of
    character substitutions necessary to transform one string into another.
  \item \textbf{Damerau-Levenshtein distance} is a classic edit-based
    algorithm, adding to the \textit{Levenshtein distance}. More
    lenient than Hamming distance it allows for a wider range of
    character manipulations, including insertions, deletions and
    transpositions of adjacent characters.
\end{itemize}

In the testing process, Python's \textit{jellyfish} library was used
for specific implementations of these algorithms.
Additionally, in order to gain the broadest perspective, similarity
comparisons were performed in two ways. First compared whole files of
original code to whole duplicate files. Second focused on generated
fragments only, weighing suggestions returned by Tabby against the
snippets from original files with the corresponding locations.

\section{Results}

\subsubsection{Static metrics outcomes}

\begin{figure}[!htb]
  \centering

  \begin{subfigure}[t]{0.49\textwidth}
    \centering
    \includegraphics[width=\textwidth]{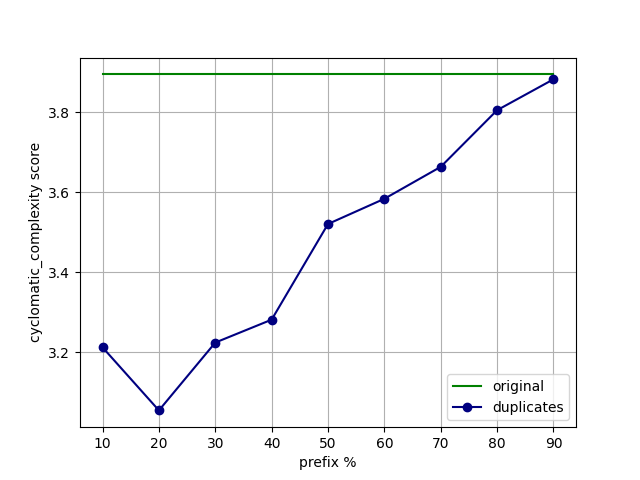}
    \caption{Averaged cyclomatic complexity.}
    \label{subfig:cc_complexity}
  \end{subfigure}%
  \begin{subfigure}[t]{0.49\textwidth}
    \centering
    \includegraphics[width=\textwidth]{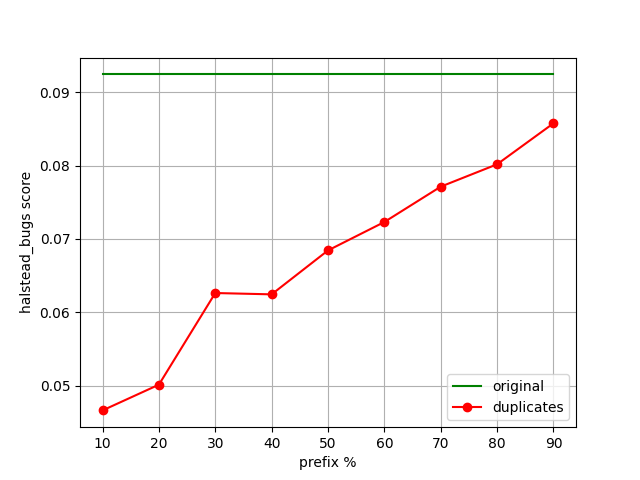}
    \caption{Averaged Halstead bugs.}
    \label{subfig:halstead_bugs}
  \end{subfigure}%
  \hfill
  \begin{subfigure}[t]{0.49\textwidth}
    \centering
    \includegraphics[width=\textwidth]{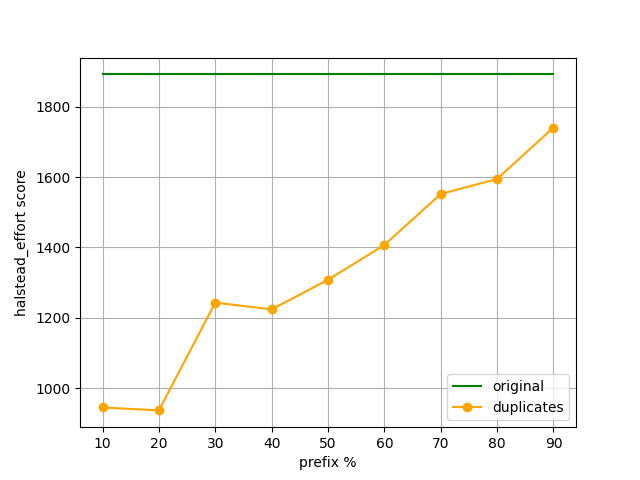}
    \caption{Averaged Halstead effort.}
    \label{subfig:halstead_effort}
  \end{subfigure}%
  \begin{subfigure}[t]{0.49\textwidth}
    \centering
    \includegraphics[width=\textwidth]{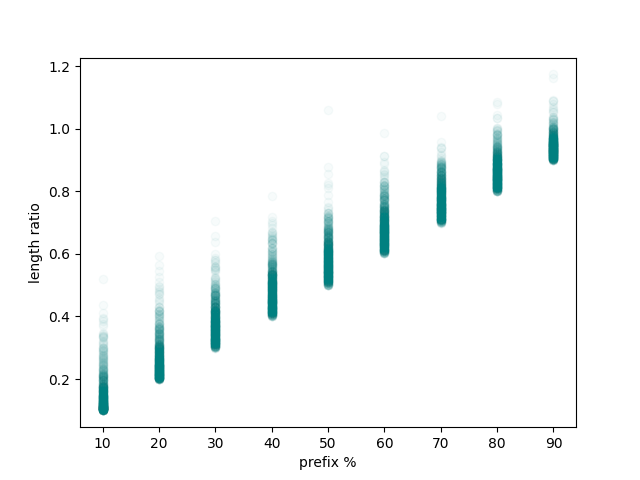}
    \caption{Length ratio of duplicate to original files.}
    \label{subfig:len_ratio}
  \end{subfigure}%

  \caption[Static metric outcomes plots]{Averaged static quality
    evaluation outcomes.\\
    Green line in plots (a), (b) and (c) denotes the averaged score of
  reference files per metric.}
  \label{fig:static_plots}
\end{figure}

The specific choice of Radon, for calculations presented a unique
challenge in evaluating non-executable code samples. Since Radon
failed to generate scores for theses cases, a workaround was
implemented: assigning the original file's score to affected samples.
While this heuristic may have slightly inflated some results, it
prevented artificially lowered scores and gaps in analysis,
preserving data integrity.
The findings---displayed in figure \ref{fig:static_plots}---confirmed
that metrics for Tabby-generated code correlate with its length,
indicating an overall healthy and stable pattern. Additionally, the
necessity of approximating scores for incomplete code reflects the
practical challenges of AI-generated code evaluation, as real-world
scenario of a workflow powered by coding assistant would be provided
with short, digestive, incremental suggestions, often not yet in an
executable form.

\subsubsection{Similarity detection}

Overall, similarity analysis displays a more consistent and focalized
behavior in scenarios of whole-file evaluations than in generated
code fragments only. SequenceMatcher and Jaro-Winkler produce
relatively high similarity scores for entire files, while edit-based
algorithms yield values around ~1000---expected given the dataset's
file size variance (~220 to ~25k characters, with an average of ~3k).

However, this consistency in whole-file metrics outcomes is largely
due to the unchanged portions of the original programs---the prefix
prompts---dominating the similarity scores. Generated fragments alone
exhibit more scattered spectrum of outcomes across all prefix ratios,
contrasting with the directionality of whole-file outcomes.

\paragraph{Jaro-Winkler Behavior \& Prefix Orientation}

Unlike SequenceMatcher that follows a close to linear manner in terms
of whole file's comparison, Jaro-Winkler features a noticeable dip at
10 and 20\% prefix ratios, before establishing a more stable score
for the middle prefixes. Despite the score being relatively
high---around 0.65-0.7 similarity---it is a visible disruption to the
general trend of gradual rise in similarity scores for the increased
amounts of context.
This can be attributed to Jaro-Winkler's bias toward early text
matches, meaning that for shorter prefix lengths, there is less of
the original program's coverage, disproportionately lowering the
scores in comparison to the following sharp increase as more of the
original content occupies the starting portion of Tabby-completed
files with bigger prefix ratios.

In the analysis of generated fragments, Jaro-Winkler also performs
notably well, with a solid representation of scores nearing 0.8-0.9.
This is due to its exceptional suitability for shorter strings, that
Tabby-generated suggestions fall into---average snippet length
typically less than 100 characters.
Nevertheless, these outcomes contribute to the positive takeaways
from the experiment and suggest potential for Tabby-generated suggestions.

\begin{figure}[!h]
  \centering
  \begin{subfigure}[t]{0.49\textwidth}
    \centering
    \includegraphics[width=\textwidth]{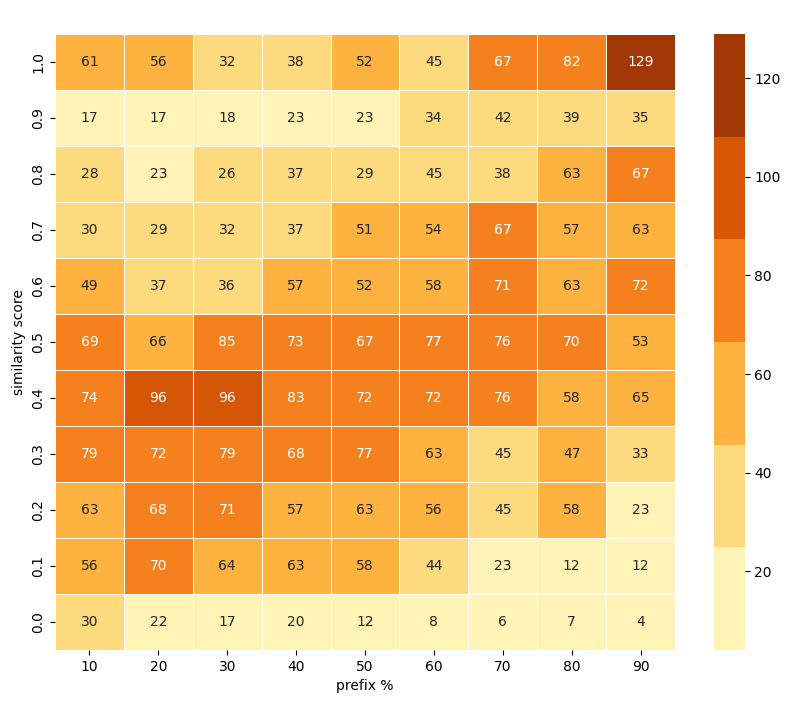}
    \caption[SequenceMatcher generated fragments comparison
    heatmap.]{SequenceMatcher generated fragments comparison heatmap.}
    \label{subfig:sequence_matcher_heatmap}
  \end{subfigure}%
  \begin{subfigure}[t]{0.49\textwidth}
    \centering
    \includegraphics[width=\textwidth]{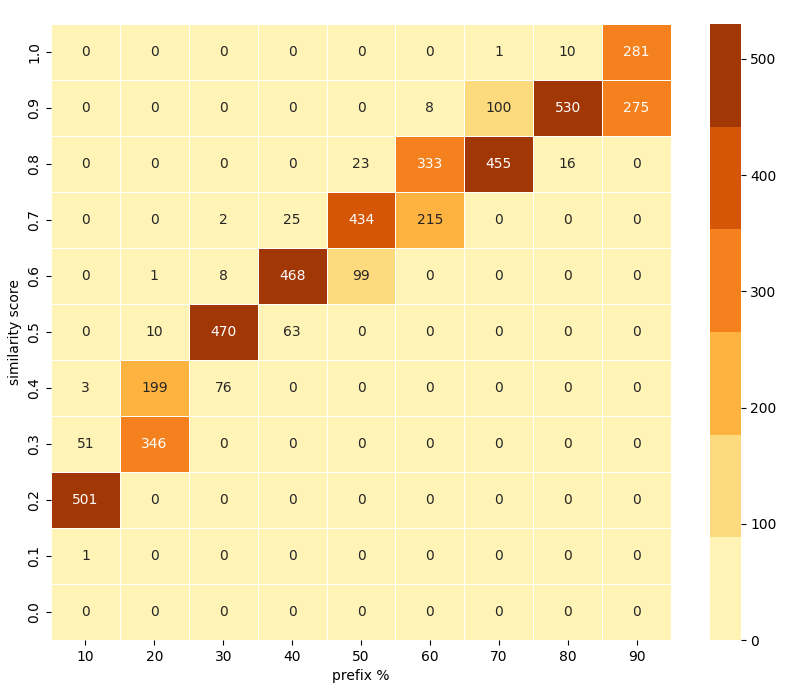}
    \caption[SequenceMatcher full files comparison
    heatmap.]{SequenceMatcher full files heatmap.}
    \label{subfig:sequence_matcher_full_heatmap}
  \end{subfigure}%
  \hfill
  \begin{subfigure}[t]{0.49\textwidth}
    \centering
    \includegraphics[width=\textwidth]{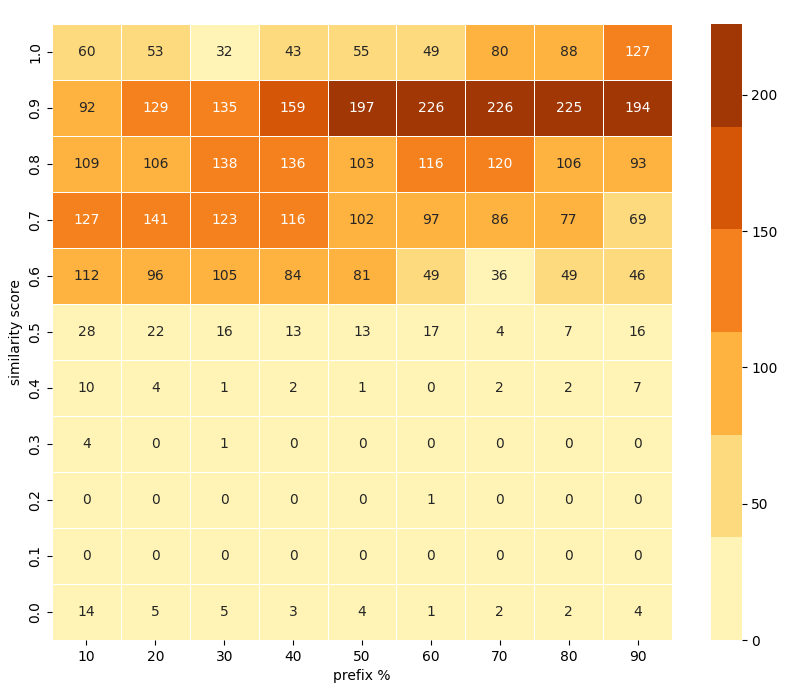}
    \caption[Jaro-Winkler generated fragments comparison
    heatmap.]{Jaro-Winkler generated fragments comparison heatmap.}
    \label{subfig:jaro_winkler_heatmap}
  \end{subfigure}%
  \begin{subfigure}[t]{0.49\textwidth}
    \centering
    \includegraphics[width=\textwidth]{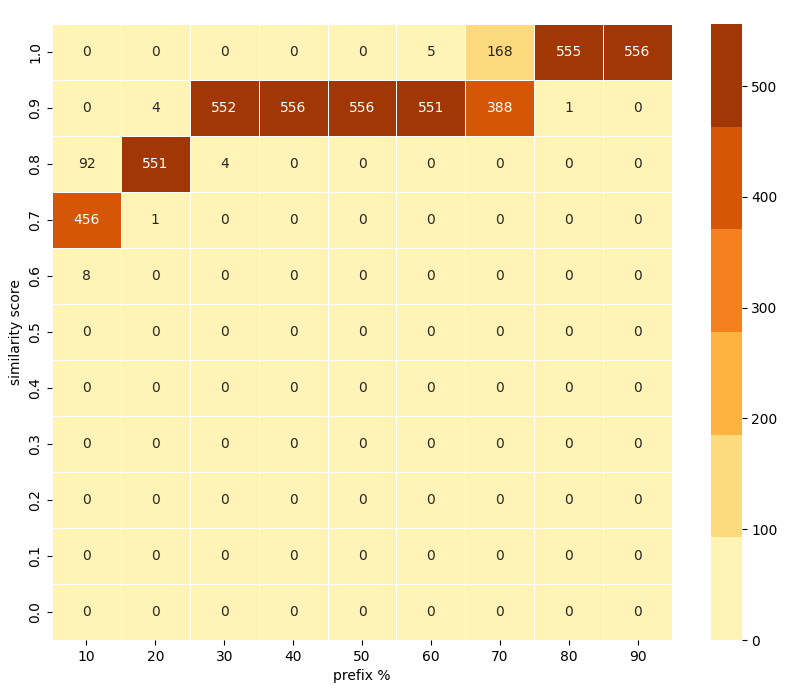}
    \caption[Jaro-Winkler full files comparison
    heatmap.]{Jaro-Winkler full files heatmap.}
    \label{subfig:jaro_winkler_full_heatmap}
  \end{subfigure}%

  \caption[SequenceMatcher and Jaro-Winkler heatmaps.]{Heatmaps for
    SequenceMatcher and Jaro-Winkler similarity algorithms outcomes,
    testing either on the whole file or on the generated snippet only.
  Outcomes in the [0, 1] range.}
  \label{fig:sequence_matcher_jaro_winkler_heatmaps}
\end{figure}

\paragraph{SequenceMatcher's Order Sensitivity}

Unlike Jaro-Winkler, SequenceMatcher is particularly sensitive to the
order of matching sequences, making it more aligned with real-world
coding where sequence greatly impacts functionality. This
characteristic explains its more scattered similarity results for
generated fragments. While Jaro-Winkler's results may appear overly
optimistic, SequenceMatcher holds a reputation of an algorithm that
is especially well suited for reflecting human intuition in
evaluating the sense of similarity.

\begin{figure}[!htb]
  \centering
  \begin{subfigure}[t]{0.49\textwidth}
    \centering
    \includegraphics[width=\textwidth]{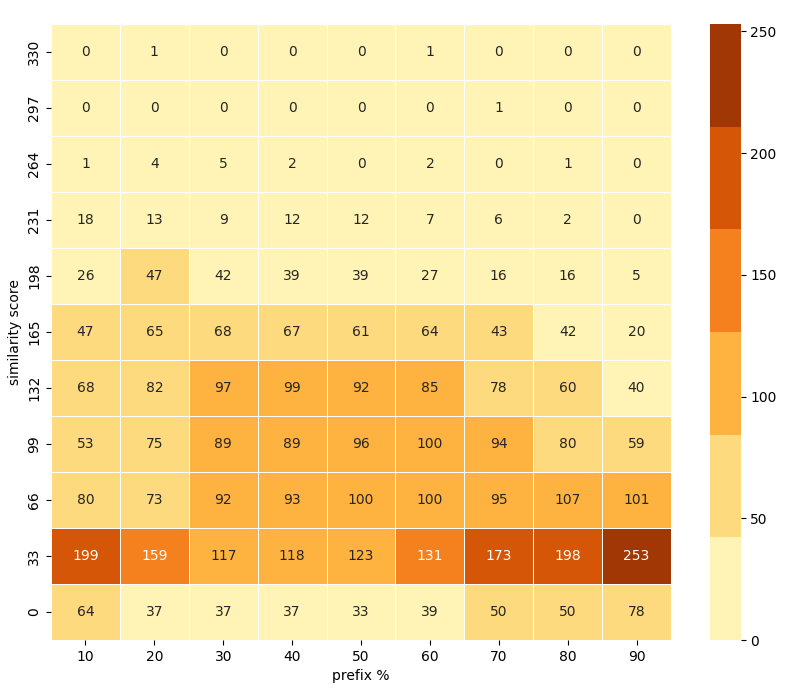}
    \caption[Damerau-Levenshtein distance generated fragments
    comparison heatmap.]{Damerau-Levenshtein generated fragments
    comparison distance heatmap.}
    \label{subfig:damerau_levenshtein_heatmap}
  \end{subfigure}%
  \begin{subfigure}[t]{0.49\textwidth}
    \centering
    \includegraphics[width=\textwidth]{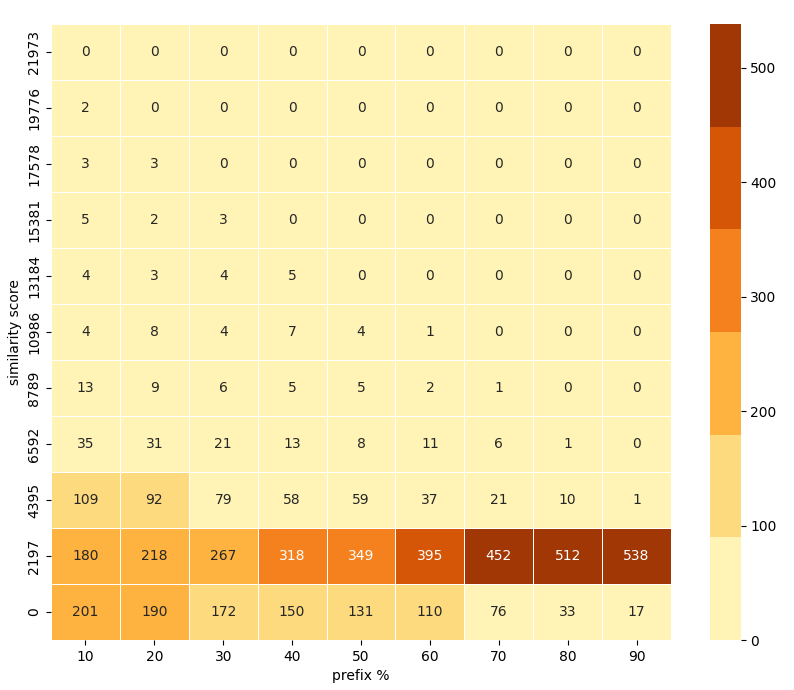}
    \caption[Damerau-Levenshtein distance full files comparison
    heatmap.]{Damerau-Levenshtein distance full files heatmap.}
    \label{subfig:damerau_levenshtein_full_heatmap}
  \end{subfigure}%
  \hfill
  \begin{subfigure}[t]{0.49\textwidth}
    \centering
    \includegraphics[width=\textwidth]{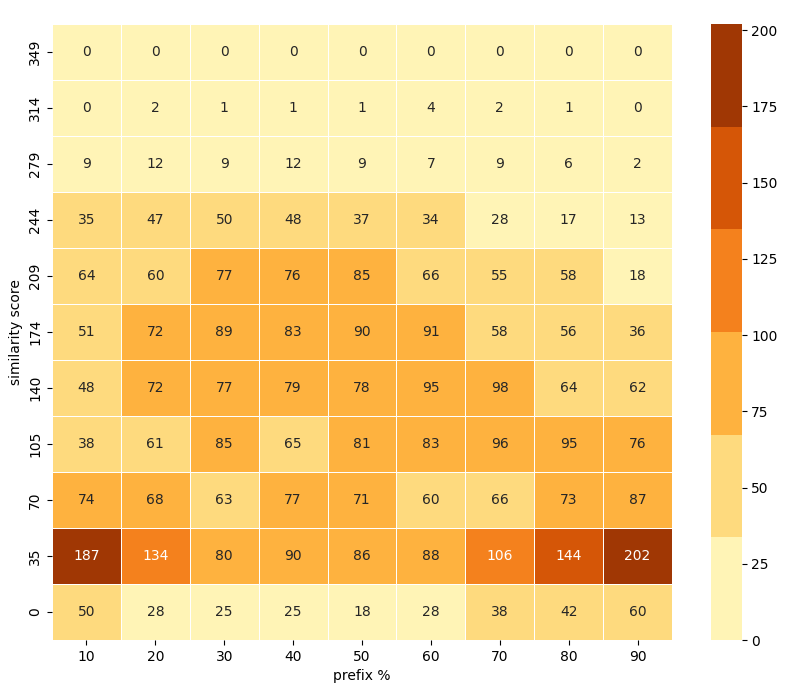}
    \caption[Hamming distance generated fragments comparison
    heatmap.]{Hamming distance generated fragments comparison heatmap.}
    \label{subfig:hamming_heatmap}
  \end{subfigure}%
  \begin{subfigure}[t]{0.49\textwidth}
    \centering
    \includegraphics[width=\textwidth]{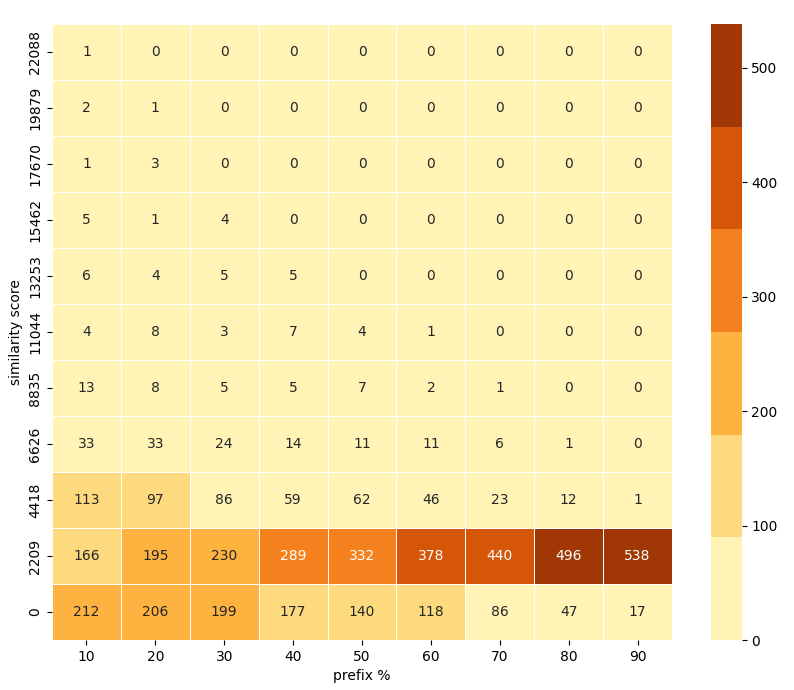}
    \caption[Hamming distance full files comparison heatmap.]{Hamming
    distance full files heatmap.}
    \label{subfig:hamming_full_heatmap}
  \end{subfigure}%

  \caption[Damerau-Levenshtein distance and Hamming distance
  heatmaps.]{Heatmaps for Damerau-Levenshtein and Hamming distance
    algorithms outcomes, testing either on the whole file or on the
  generated snippet only. Outcomes in the [0, 1] range.}
  \label{fig:damerau_levenshtein_hamming_heatmaps}
\end{figure}

\paragraph{Distance-based Metrics: Hamming vs. Damerau-Levenshtein}

The comparison of Hamming distance and Damerau-Levenshtein reveals a
broad score distribution, reflecting varied suggestion quality. As
expected, Hamming consistently yields higher scores due to its
stricter character manipulation rules. In contrast,
Damerau-Levenshtein with its more lenient allowance for insertions,
deletions and transpositions is better suited for capturing broader
similarity trends.

However, neither of these two algorithms is particularly well fitted
for capturing the broad sense of similarity, that SequenceMatcher and
Jaro-Winkler normalized in the (0, 1) range provide. Their role is to
provide a different kind of perspective, one where the numerical
outcomes are highly specific to the lengths of the strings considered.

Nevertheless, with a solid representations of outcomes nearing the
value of 0, they build upon the findings of two previous algorithms.
All of the metrics defined for this testing process present a
satisfactory number of values in acceptable ranges.

These findings positively support Tabby's ability to provide accurate
suggestions, closely matching certain scenarios and encourage further
exploration of this coding assistant.

\section{Conclusions}

While this study demonstrates promising results for code generated by
Tabby, signifying its value as a coding assistant, several
limitations in the testing process must be acknowledged. The reliance
on string-based similarity detection, rather than a clone detection
technique able to detect functional clones, proves to be the most
inherent challenge in evaluating true code quality. This choice most
likely resulted in the lowered outcomes of the testing process.
Additionally, the selection of prefixes for the prompting could be
enhanced. Even though the increasing percentage steps method provides
enough randomization, prefix selection based on approach such as
abstract syntax tree (AST) could result in broader insights.
Another limitation lies in the choice of model, as testing was
conducted exclusively with the smallest available LLM from Tabby's
repository, without comparisons with performance of the more complex models.
However, despite these concerns, the bottom line of the study suggest
that Tabby is a promising tool for code generation, worth of further
exploration.

More importantly, this study's underscores the need for continuous
evaluation of generative AI tools to ensure that they meet evolving
developer needs.

\subsubsection*{Author contributions}

M.B and R.N. identified the problem, M.B. designed the approach, downloaded the data, implemented the software, performed numerical experiments, M.B. and R.N. prepared the draft. All authors have read and agreed to the published version of the manuscript.

\subsubsection*{Funding}

A statutory Research Grant from the Institute of Computer Science, Warsaw University of Technology, supports this work.

\subsubsection*{Software availablity}

The presented pipeline is available on the
\url{https://github.com/metredecoeur/tabby-testing-pipeline}
repository under the MIT licence.

\subsubsection*{The authors declare no conflict of interest.}

\end{document}